\documentclass[aps,twocolumn,showpacs,floats,superscriptaddress]{revtex4}
\usepackage{graphicx}
\usepackage{dcolumn}
\usepackage{bm}
\def\he4{$^4$He}
\def\Am3{\AA$^{-3}$}
\def\beq{\begin{equation}}
\def\eeq{\end{equation}}
\begin{document}

%%%%%%%%%%%%%%%%%%%%%%%%%%%%%%%%%%%%%% AUTHORS %%%%%%%%%%%%%%%%%%%%%%%%%
\author{M. Boninsegni}
\affiliation{Department of Physics, University of Alberta,
Edmonton, Alberta T6G 2J1}

\author{A.B. Kuklov}
\affiliation{Department of Engineering Science and Physics, CUNY,
Staten Island, NY 10314}

\author{L. Pollet}
\affiliation{Theoretische Physik, ETH Z\"urich, CH-8093 Z\"urich,
Switzerland}

\author{N.V. Prokof'ev}
\affiliation{Department of Physics, University of Massachusetts,
Amherst, MA 01003, USA}
 \affiliation{Russian Research Center
``Kurchatov Institute'', 123182 Moscow, Russia}

\author{B.V. Svistunov}
\affiliation{Department of Physics, University of Massachusetts,
Amherst, MA 01003, USA} \affiliation{Russian Research Center
``Kurchatov Institute'', 123182 Moscow, Russia}

\author{M. Troyer}
\affiliation{Theoretische Physik, ETH Z\"urich, CH-8093 Z\"urich,
Switzerland}
%%%%%%%%%%%%%%%%%%%%%%%%%%%%%%%%%%%%%%%%%%%%%%%%%%%%%%%%%%%%%%%%%%%%%%%%%%%%%%

\title{Luttinger Liquid in the Core of Screw Dislocation in
Helium-4}

%%%%%%%%%%%%%%%%%%%%%%%%%%%%%%%%%%%%%%%%%%%%%%%%%%%%%%%%%%%%%%%%%%%%%%%%%%%%%%
\date{\today}
\begin{abstract}
On the basis of first-principle Monte Carlo simulations  we find
that the screw dislocation along the hexagonal axis of an {\it hcp}
\he4 crystal features a superfluid (at $T\to 0$) core. This is the
first example of a regular quasi-one-dimensional supersolid -- the phase featuring both translational and superfluid orders, and
one of the cleanest cases of a Luttinger-liquid system. In
contrast, the same type of screw dislocation in solid H$_2$ is insulating.

\end{abstract}

 \maketitle

The remarkable observations of a non-classical moment of inertia
in solid \he4 \cite{KC,Rittner1,Shirahama,Kubota,Rittner2} have
sparked a wave of interest in the supersolid state of matter (see
review \cite{Prokofev07} and references therein). According to the
standard paradigm {\cite{andreev69}, the uniform supersolid state
develops if the energy gap for creating vacancies vanishes in the
ground state of a crystal, leading to a finite concentration of
stable and delocalized vacancies. In the past, however,
experiments did not find such zero-point vacancies
\cite{novacancies}. Recent Monte Carlo simulations of solid \he4
{\cite{noSFSus,Bernu,superglass,Clark} have further limited
possible mechanisms for a uniform supersolid state:  A perfect
{\it hcp} \he4 crystal is an insulator with a large energy gap
toward vacancy creation and an even larger one for interstitials.
When vacancies are introduced by hand into a perfect {\it hcp}
crystal, they undergo phase separation \cite{noSFSus}. This 
provides strong evidence that a \he4 crystal does not conform
to any standard supersolid scenario.

The situation is quite different for defect-ridden, or polycrystalline {\it hcp}
\he4.
Simulations showed that grain boundaries are generically
superfluid \cite{GB_SF}, while a recent experiment observed
superflow in non-uniform samples and attributed it to the presence
of grain boundaries  {\cite{Balibar}}. 
In general, two types of supersolid may exist -- regular and glassy. 
While in both cases translational symmetry is broken, regular supersolid is
characterized by two orders -- discrete translational and superfluid.
Thus, due to lack of discrete translational order,
the superfluid grain boundaries do not constitute an example of
a regular supersolid. They should rather be considered as superglass.

%Grain boundary
%superfluidity (SF) is promoted by strong topological frustration at the
%interface between microcrystals. Importance of the
%degree of frustration is clear from the observation that grain
%boundaries with zero twist angle and small tilting angles remain
%insulating \cite{GB_SF}. In general, due to the lack of translational order,
%the superfluid grain boundaries do not constitute an example of
%supersolid.

One may conjecture that (straight) dislocations, featuring
translational order, might host SF along their cores
\cite{Shevchenko, Meyerovich}. More recently, a phenomenological
model was proposed, where crystal deformations~\cite{Dorsey}, 
particularly inside a dislocation core \cite{Dorsey2}, promote
SF see also Ref. \cite{monas} on role of $^3$He in inducing superfluidity in dislocation cores. None of the previous theories, however, was quantitative and conclusive. 
The model \cite{Dorsey,Dorsey2} contrasts with our observation of
insulating grain boundaries with small tilt angles because
this type of boundary can be considered as a wall of rarely spaced {\it edge}
dislocations \cite{GB_SF, Landaushitz}. 
We are not aware of any suggestion that screw dislocations 
have any advantage over edge dilocations in terms of their SF 
properties; on the contrary, they were often excluded from the list. 
%In general, we find
%that defects with low degree of disorder are insulating. % \cite{note}.
Thus, the question remains: Do extended supersolid defects 
characterised by translational and superfluid orders exist in
\he4 crystals?
In this Letter, we give an affirmative answer to this question, i.e., we
provide theoretical evidence of SF inside the core of a screw dislocation aligned with
 the $c-$symmetry axis of {\it hcp} \he4.

Our grand-canonical Monte Carlo simulations are based on the worm
algorithm, i.e., a Path Integral Monte Carlo
technique defined on the configuration space for the
single-particle Matsubara Green function \cite{worm}. For spatial imaging,
we employ the winding-cycle maps, i.e., averaged set of points measuring 
instant spatial position of a randomly selected element (bead) of the worldline
with non-zero winding number (winding cycle). The winding cycles
define the superfluid response \cite{Pollock}, so that the
winding-cycle map allows one to visualize the spatial distribution
of the superfluid component. Qualitatively speaking, a winding-cycle map 
represents the local superfluid response.

{\it Sample geometry.} The geometry of our sample is illustrated
in Figs.~\ref{fig1} and \ref{fig2}. The sample is periodic in the
$z$-direction, with 8 pairs of basal planes. In order to rule out 
finite-size effects, we have
simulated samples with 3 and 4 pairs of basal planes as well.

Cell periodicity in the $xy$ plane is  incompatible with the 
presence of a single dislocation. One could restore it 
by simulating a dislocation pair, but we opt instead 
for a single dislocation, using a non-periodic cell 
in the $xy$ plane, as this allows us to {\it a}) simulate samples 
with the rotational symmetry of a {\it hcp} crystal {\it b}) obtain 
robust results with smaller samples.

Our simulation sample consists of an ideal
{\it hcp} crystal with a screw dislocation in its center along the
$c$ axis. Specifically, we, first, superimpose 
two identical triangular layers 
separated by half of the $c$-axis period $a_z=\sqrt{8/3}a$ along
the $c$-axis and shifted by $(a/2, \, a/\sqrt{3})$ in the
$xy$-plane, $a$ being the lattice period along the $x$-axis.
The crystal, then, has been cut along the $xz$-half-plane, whose edge
(located at the center of one of the hexagons formed when all basal layers are projected to a single $xy$-plane)
 is the dislocation core. Atoms have been  
displaced so that, upon completing a full revolution around the
core, each atom advances by a lattice period 
$a_z$. This procedure creates an ideal (classical) screw dislocation in essentially infinite medium. 
Then, all particles located outside a (pencil shape) cylinder, with the dislocation core being its axis of symmetry, (and inside the rectangular simulation cell) have been pinned. These particles are not moved in our 
Monte Carlo simulation, but they do interact with the rest of the system via
the helium pair potential. The purpose of these ``frozen" particles is
that of confining our sample to the inner (cylindrical) part of the cell, while, practically, completely eliminating 
the effects of open boundary conditions. 
Our rectangular simulation cell is twice the size of the  physical
sample inside the cylinder. The initial numbers of (physical) atoms used in simulations are $N_{ini} = 384,\,   768,\,  1912$.

%%%%%%%%%%%%%%%%%%%%%%%%%%%%%%%%%%%%%%%%%%%%%%%%%%%%%%%%%%%%%%%%%%
\begin{figure}[t]
\centerline{\includegraphics[bb = 0 300 600 350, scale=0.425,
angle=-90]{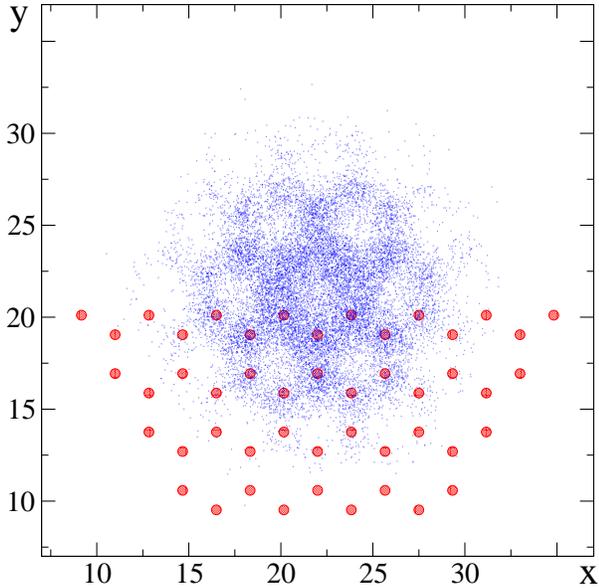}} \caption{\small{(Color online. For the best
perception, look from a distance.) Columnar winding-cycle map
(blue dots) in the core of the screw dislocation in solid \he4 at
$T=0.25\,$K and density 0.0287$\,$\AA$^{-3}$. View is along the
hexagonal axis, with the core parallel to it. Shown with red dots
(in the lower half of the plot only) are the atomic positions in
the initial configuration. At distances $\sim 3a$ from the core,
the atomic positions averaged over the imaginary time are only
slightly shifted against those of the original configuration.The
unit of length is 1 \AA. }}\label{fig1}
\end{figure}
%%%%%%%%%%%%%%%%%%%%%%%%%%%%%%%%%%%%%%%%%%%%%%%%%%%%%%%%%%%%%%%%%%%%%%%%
\begin{figure}[t]
%\centerline{\includegraphics[bb = 10 250 600 350, scale=0.473,
\centerline{\includegraphics[bb = 10 200 600 350, scale=0.473,
angle=-90]{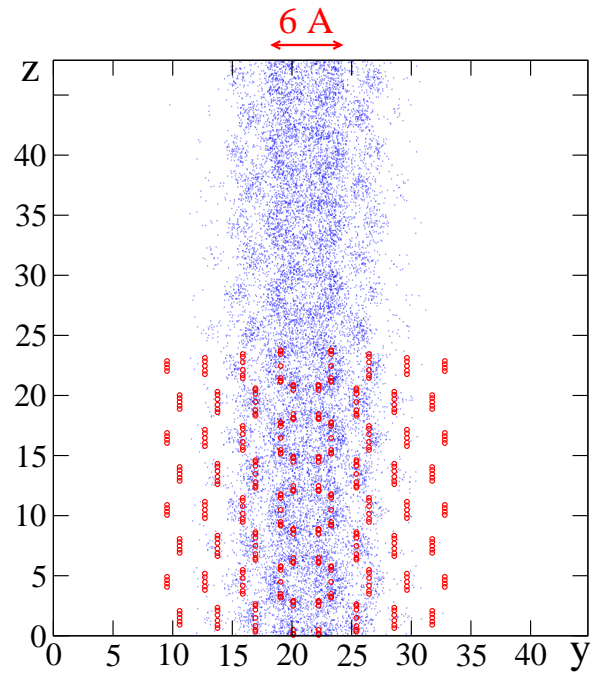}} \caption{\small{(Color online.)  Columnar
winding-circle map  in the core of screw dislocation. View is
along the $x$-axis in the basal plane---perpendicular to the core.
Shown with red dots (in the lower half of the plot only) are the
atomic positions in initial configuration. The unit of length is
1 \AA.}}\label{fig2}
\end{figure}
%%%%%%%%%%%%%%%%%%%%%%%%%%%%%%%%%%%%%%%%%%%%%%%%%%%%%%%%%%%%%%%%%%%%%%%%%%%%%%%%%%%%%%%%%%%%%%%%%%%%%%%%%

{\it Superfluid response.} The simulations were performed at the
number density 0.0287$\,$\AA$^{-3}$, that is, in the close vicinity
of the melting point. The lattice period along the $x$-direction
is then $a=3.666\,$\AA. The convenience of working at the
melting point is that we know the chemical potential, $\mu =
0.02\,$K, from the simulations of the liquid phase at its freezing
density \cite{worm}. Our simulations span a temperature range 
from 0.2 to 1 K.

In Figs.~\ref{fig1} and \ref{fig2} we present the projected
winding-cycle maps on the $xy$- and $yz$-planes, together with the initial configuration of the atoms.
We see that the superfluid density in the dislocation core
is strongly correlated with the insulating environment, respecting
the hexagonal symmetry modified by the presence of the screw dislocation.
The superfluid density is most robust along the bonds of
the ideal structure.

Quantitatively, our data can be analyzed using the concepts of
Luttinger liquid theory. Luttinger liquids are characterized by
two parameters, the superfluid stiffness $\Lambda_s=n_s^{(1D)}/m$
(where $n_s^{(1D)}$ is the 1D superfluid density and $m$ is the
helium atom mass)
 and the compressibility $\kappa$. Both can be
extracted from the distributions $P_W(W), \, P_N(N)$ of the
winding $W$ (along the core) and particle number $N$ fluctuations
directly obtained in the simulation:
\begin{eqnarray}
P_W(W) & \sim & \exp \left( - \frac{ L  W^2} {2 \beta \Lambda_s } \right)\, , \nonumber \\
P_N(N) & \sim & \exp \left( - \frac{\beta N^2}{2  L \kappa}
\right)\, ,
\end{eqnarray}
where $N$ is taken relative to its mean $\langle N \rangle$
\cite{worm1}, $L$ is the sample length along the $c$-axis and
$\beta$ stands for inverse temperature. We fitted the collected histograms $P_W(W), \, P_N(N)$ with Gaussians and have obtained the superfluid
stiffness and the compressibility. The Luttinger parameter is
given as $K_L^{-1} = \pi \sqrt {\Lambda \kappa}$. Our data (for the
different temperatures, and different system sizes in the $z$-direction)
yield $K_L=0.205(20)$, where the finite-size (along the $c$-axis)
and statistical errors are combined. The initial number of
(updatable) particles in the state shown in Figs. \ref{fig1} and
\ref{fig2} was $N=768$ corresponding to a relatively tight
confinement in the $xy$-plane. We clearly see superfluidity in
samples with much larger number of particles in the $xy$-plane
(1912 updatable particles in the sample with 3 pairs of basal
planes). In particular, the superfluid stiffness was found to be
the same within the error bars (of about 20\%). However, the
accuracy of data for $\kappa$ was not sufficient for resolving
finite-size effects in $K_L$ due to the $xy$-confinement.

For the value of the chemical potential corresponding to the melting point, the
average equilibrium number of particles in the 8 basal-pair sample
was found to be about 770. By varying chemical potential and, in
particular, lowering it below the point where $N=768$, 
we saw no qualitative
difference in the superfluid response, which is fully
consistent with the fact that $K_L<1/2$ (in the Luttinger liquid,
an external periodic potential is relevant for opening the
insulating gap only at $K_L>1/2$
\cite{Coleman,Chui,Haldane,Giamarchi}). A
sample half the size in $z$-direction yielded the same
qualitative conclusion. In the 3 basal-pair sample with much
larger number of particles in the $xy$-plane, we observe that the
core dopes itself with vacancies (of the order of 3-4), rather than with
interstitial atoms, apparently
due to lower effective chemical potential in the core. We do not
attribute special significance to this fact because by increasing
the chemical potential by $4~K$, far smaller than the interstitial
energy in the {\it hcp} solid, the system can be made to have the
original (classical solid) particle number without changing the
superfluid nature of the state.

From our data for $\Lambda_s$ we deduce that $n_s^{(1D)}$ is about
1 particle per 1$\,$\AA, which is equivalent to saying that  superfluid
phase in the dislocation core involves nearly all atoms in a ``tube" of diameter
6 \AA. More quantitatively, one can
assign the superfluid response to partial contributions from the
coordination shells. Nearly $90\,$\% of points in the
winding-cycle map in Fig.~\ref{fig1} belong to the first three
coordination shells with the following division: $35\,$\% in the
first shell, $29\,$\% in the second one, and $26\,$\% in the third
one.

Special care should be taken to make sure that the observed
superfluid response is not an artifact of poor thermalization of
the worldline configurations in the dislocation core. This concern
stems from our previous experience with edge dislocations, where we
saw that a superfluid glassy region in the core---created either
by hand, or as a result of quenched relaxation of the initial
configuration---remains essentially metastable, at least at
temperatures $\sim 0.2\,$K. To rule out such a scenario, we first relaxed
the initial configuration by treating  atoms as classical particles with a
repulsive potential $\sim 1/r^6$, and minimizing the energy of the
configuration. [In the case of edge dislocations, this protocol
leads to an insulating groundstate.] Our observation was that in
both cases---with or without classical relaxation---the Monte
Carlo process converges to one and the same result. Another
indication that our superfluid signal is not an artifact of poor
thermalization comes from the perfect regularity of the superfluid
density map, in combination with its strong correlation to the
insulating environment. This outcome rules out concerns based on a
lack of thermalization in the glassy state.
Analogous simulations have been performed for the {\it hcp}
{\it para}-hydrogen
crystal, using the same sample preparation procedure, at a density
of 0.0261 \AA$^{-3}$, corresponding to the $T\to 0$ equilibrium density. 
In this case, the dislocation core is found to be an insulator
{\it doped with vacancies} (forming a commensurate density wave).

{\it Discussion}. The observation of a superfluid dislocation in a
\he4 crystal is an interesting  example of a regular
quasi-one-dimensional supersolid. However, the question 
%which naturally arises 
is whether this observation is relevant to the
effect of non-classical inertia
% discovered by Kim and Chan
\cite{KC}. 
%Clearly, 
Superfluid dislocations can be a part of the
disordered superfluid network, together with superfluid grain
boundaries, ridges, and liquid (or glassy, at higher pressure)
pockets, etc. Considering a network of dislocations only, 
%one has to keep in mind that 
the effective three-dimensional superfluid
fraction of such a network is limited by the volume fraction of
the atoms located at the cores of dislocations. This fraction is
likely to be significantly smaller than $\sim 1\,$\%. The
dislocation density in solid \he4 typically varies over a range as wide as
$10^6\div 10^{10}\,$cm$^{-2}$ \cite{dens1,dens2,dens3,dens4,Kosevich}.
%, with the controlling mechanisms being the crystal growth near an non-wetted wall \cite{Kosevich} as well as the dislocation motion and recombination during the annealing \cite{dens4}. 
It is, however, also possible
to grow crystal without screw dislocations at all \cite{zeroscrew}. Taking the 1D
superfluid density of $n_s^{(1D)} \approx 1 \,$\AA$^{-1}$ observed
in our simulations, we estimate the required order-of-magnitude
dislocation density $n_d\, \sim\, 1/l^2 \, \sim\,  3\times
10^{12}\,$cm$^{-2}$ in order to account for $n_s^{(3D)}/n\, =\,
(n_s^{(1D)}n_d) /n \, \sim\,  1\,$\% of the superfluid fraction
observed in \cite{KC}. Here $l$ is the typical distance between
dislocations. 
%This is at least two orders of magnitude larger than the expected upper value of the dislocation density.

The Luttinger liquid parameter is all we need to characterize
superfluid properties of the dislocation network consisting of
segments of length $l\gg a$. As it has been pointed out by
Shevchenko \cite{Shevchenko2}, in this system one has to
distinguish between  static and dynamic responses. Below the
thermodynamic transition temperature $T_c \sim T_* a/l~$ (which
can be also written as $T_c\, \sim\,  T_*\, [n_s^{(3D)}/n]^{1/2}$,
in terms of the zero-point superfluid fraction $n_s^{(3D)}/n$) the
network features a non-zero static superfluid response
\cite{Shevchenko2}. Here, $T_*$ is the characteristic temperature
of the superfluid helium liquid. In our simulations we find that
the core has robust phase coherence properties at all temperatures
below $1\,$K, which allows us to set $T_* \sim 1\,$K (the
continuation of the $\lambda$-line in liquid helium to the region
of solid densities gives $T_\lambda\approx 1.5\,$K \cite{GB_SF}).
This estimate is in agreement with the small value of $K_L$.

In the temperature range $T_c < T < T_*$, which is very broad for
large aspect ratio $l/a$, the state of the system can be viewed as
a vortex tangle with `vortex cores' (which are not well defined
geometric objects because of the natural uncertainty $\sim l$ in
their sizes and positions) pinned in the solid bulk. We refer to
this state as Shevchenko state \cite{Shevchenko2}. In contrast to
the original purely classical analysis of Ref.~\cite{Shevchenko2},
we emphasize that the dynamics of Shevchenko state should be
driven by discrete {\it quantum-tunneling} phase slip events
(instantons), with the typical relaxation time $\tau \sim
(T_*/T)^{8/K_L-1}/T_*$, as it follows from the microscopic
analysis \cite{95}. In the low-frequency limit $\omega \tau \ll 1$
the state is essentially normal. At ``high'' frequencies $\omega
\tau \gg 1$ the response is nearly indistinguishable from the
standard superfluid \cite{Shevchenko2}. Note also, that for $K_L
\ll 1$ and $T \ll T_*$ the phase-slip time can easily exceed any
realistic experimental time-scale making it very hard to detect
the genuine thermodynamic transition at $T_c$.

While being focused on one isolated screw
dislocation, we note that small
twist-angle grain boundary can be represented as an array of screw
dislocations \cite{Landaushitz} and, therefore, must also be
superfluid with an additional feature of tunneling coupling
between the cores. Such boundaries might be the most realistic objects
for studies of superfluidity of screw dislocations.

In summary, we have found that the screw dislocation in {\it hcp}
solid \he4 supports (at $T\to 0$)  superfluid transport of \he4
atoms along its core, yielding thus the first realistic example of
a regular continuous supersolid. Analogous dislocation in the {\it
hcp} solid H$_2$ is found to be an insulator.

The authors acknowledge valuable discussions with Alex Meyerovich,
Moses Chan, Miko Palaanen, Sergei Shevchenko and Sebastien Balibar. This work was
supported by the Swiss National Fund, the National Science Foundation under Grants Nos.
PHY-0426881, PHY-0426814 and PHY-0456261, and by the Natural Science and
Engineering Research Council of Canada under grant  G12120893. We recognize an
essential role of  (super)computer clusters at UMass, Typhon at
CSI, Hreidar at ETH, and AICT, University of Alberta. N. P. and B. S. acknowledge partial
support from CRDF under Grant 2853. A.K. acknowledges partial support from
PSC-CUNY Grant 68230-0037.

\end{document}